\documentclass[useAMS,usenatbib,psfig]{mn2e}

\usepackage{psfig}
\usepackage{times}

\newcommand{\bd}{\begin{displaymath}}
\newcommand{\ed}{\end{displaymath}}
\newcommand{\be}{\begin{equation}}
\newcommand{\ee}{\end{equation}}

\title[Constraints on radiatively inefficient accretion history]
{Constraints on radiatively inefficient accretion history from
Eddington ratio distribution of active galactic nuclei}
\author[X. Cao \& Y.-D. Xu]
{ Xinwu Cao$^{1}
\thanks{E-mail: cxw@shao.ac.cn}$,
Ya-Di Xu$^2
\thanks{E-mail: ydxu@sjtu.edu.cn}$\\
$^1$ Shanghai Astronomical Observatory,
Chinese Academy of Sciences, 80 Nandan Road, Shanghai, 200030, China\\
$^2$ Physics Department, Shanghai Jiaotong University, 1954
Huashan Road, Shanghai, 200030, China }

\date{Accepted 2007 February 09}

\pagerange{\pageref{firstpage}--\pageref{lastpage}} \pubyear{}

\begin{document}

\maketitle \label{firstpage}

\begin{abstract}

The transition of a standard thin disk to a radiatively
inefficient accretion flow (RIAF) is expected to occur, when
$\dot{m}\sim \dot{m}_{\rm crit}$ ($\dot{m}=\dot{M}/\dot{M}_{\rm
Edd}$). The radiative efficiencies of accretion flows accreting at
rates lower than the critical accretion rate $\dot{m}_{\rm crit}$ become significantly
lower than that of standard thin disks.
It is believed that the initial
transition radius is small just after the accretion mode
transition, and then the transition radius increases with
decreasing accretion rate, as suggested by some theoretical models and observations.
Based on such variable transition radius models, we derive how the accretion rate
$\dot{m}(t)$ evolves with time from the observed Eddington ratio distribution for
a sample of low-luminosity active galactic nuclei in the local universe. The derived time-dependent
accretion rates $\dot{m}(t)$ show a rapid decrease after the transition of the standard
thin accretion disk to a RIAF, which is consistent with that derived from the hard
X-ray background.
\end{abstract}


\section{Introduction}

Accretion onto massive black holes is believed to power active
galactic nuclei (AGNs).  The UV/optical continuum emission
observed in luminous quasars is attributed to the thermal
radiation from the accretion disks surrounding the massive black
holes in quasars \citep*[e.g.,][]{sm89}. In recent years, some
different approaches are proposed to measure the masses of the
central black holes in AGNs \citep*[e.g.,][]{p93,fm00,g00}, and
the central black hole masses of many AGNs can be measured fairly
accurately. It is found that a fraction of luminous AGNs are
accreting at extremely high rates. Their bolometric luminosities
are around (or even higher than) the Eddington luminosity, for
example, the black holes in many narrow-line Seyfert 1 galaxies
are believed to be accreting at Eddington (or super-Eddington)
rates \citep*[e.g.,][]{s00,whd04,bz04,cw04}. Besides the luminous AGNs
accreting at high Eddington rates ($\ga 0.01$), many nearby
low-luminosity AGNs are found to have similar characteristics as
those luminous AGNs, but with relatively weaker broad-line
emission and lower luminosities in different wavebands \citep{h97}.
Most of these low-luminosity AGNs are
accreting at highly sub-Eddington rates ($L_{\rm bol}/L_{\rm
Edd}\ll 0.01$). The Eddington ratios of AGNs can spread over more
than ten orders of magnitude (from $\la 10^{-10}$ to more than
unity)\citep*[e.g.,][]{wc04,hnh06}.

It is believed that standard thin disks (or slim disks) are
present in luminous AGNs, while the radiatively inefficient
accretion flows (RIAFs) are suggested to be present in those
low-luminosity AGNs accreting at relatively low rates \citep{ny94,bb99,bb04,i00,n00}.
There is a critical accretion rate $\dot{m}_{\rm crit}$, above which the RIAF is
suppressed and a standard thin disk is present.
The RIAF may connect to a standard thin disk at
a certain transition radius $R_{\rm tr}$. This is required by modelling on a
variety of observations of AGNs
\citep*[e.g.,][]{q99,lw00,c03,k04,y04}, and is also predicted by some
theoretical model calculations \citep{a95,liu99,rc00,sd02}. In all
these models, the transition radius $r_{\rm tr}$ ($r_{\rm tr}=R_{\rm tr}/R_{\rm S}$, and
$R_{\rm S}=GM_{\rm bh}/c^2$) is expected to increase with decreasing accretion rate $\dot{m}$.

The evolution of central engines in AGNs is mainly governed by
accretion processes. The black hole accretion processes are
regulated by the gases supplied near the black hole, i.e.,
sufficient gases supplied probably lead to a high accretion rate,
and vice versa \citep{sdh05,dsh05,hh06}. As the gases being swallowed
by the black hole, the accretion rate decreases with time. An AGN
will die off, while the gases near the black hole are finally
exhausted.  How the accretion rate $\dot{m}$ evolves with time remains to be an open issue,
though an exponentially time-dependent accretion rate $\dot{m}(t)$ was widely employed in
many previous works \citep*[e.g.,][]{pv90,hl98,kh00}.
\citet{mlc81} found accretion rate $\dot{m}(t)\propto t^{-2}$, if
the gases accreted by the black hole are assumed to be supplied by
stellar collisions or tidal disruptions in a dense star system
surrounding the black hole. Compared with exponentially
time-dependent accretion rate, this form of accretion rate changes
very slowly with time. Many quasar evolution model calculations
showed that the exponentially time-dependent quasar light curve
(or simply a step function quasar light curve) can well reproduce
the observed quasar luminosity functions \citep*[e.g.,][]{hl98,wl02}.
Recently, the numerical simulations on the quasar activity
triggered by the galaxy merger showed that the quasar accretion
rate curve is very complicated \citep{sdh05,dsh05}. In their
simulations, the gases near the black hole are blown away by the bright
quasar radiation, and then accretion rate declines rapidly to
switch off the quasar activity.

In principle, the evolution of the AGN light curve $L(t)$ can be derived
from the Eddington ratio distribution of an AGN sample, because the
Eddington ratio distribution $\xi[\log (L/L_{\rm Edd})]\propto
[{\rm d}\log(L/L_{\rm Edd})/{\rm d}t]^{-1}$ \citep*[e.g.,][]{bc04,hnh06}. However, it is still not
straightforward to derive the time-dependent accretion rate
$\dot{m}(t)$, because the radiative efficiency is no longer a constant
when the accretion disk is evolving from slim disk (high-$\dot{m}$) \citep*[e.g.,][]{a88,w99}
to a RIAF (low-$\dot{m}$) \citep*[e.g.,][]{ny94}.

In this work, we derive how the accretion rate $\dot{m}(t)$ evolves with time
from an observed Eddington ratio distribution for a low-luminosity AGN sample in
the local universe based on
the standard disk/RIAF transition model. In our calculations, the physics of different
accretion models and the accretion mode transition is properly considered.

\section{Spectra of radiatively inefficient accretion flows}

The transition of a standard thin disk to a RIAF
occurs, when the accretion rate $\dot m$ decreases to a value
below ${\dot m}_{\rm crit}$. The structure of the RIAF is well
described by the self-similar solution, except the region near the black hole \citep{ny95,y96}.
The spectrum of a RIAF, $L_{\lambda}(M_{\rm bh}, {\dot m}, \alpha,
\beta)$, can be calculated approximately based on the self-similar solution
\citep*[e.g.,][]{m97,ccy02,wc04}, if the parameters $M_{\rm bh}$, $\dot m$, $\alpha$, and
the fraction of the magnetic pressure $\beta$, are specified.
However, the physical quantities of a self-similar RIAF deviate significantly from the global solution
of the RIAF in its inner region, which may lead to inaccuracy for its spectral calculation,
because most gravitational energy of the accreting matter is released in the inner
region of the accretion flow.

In this work, we calculate the global structure of the accretion flows surrounding massive
Schwarzschild black holes. The accretion flow is described by a set of general relativistic
hydrodynamical equations \citep*[see][for the details]{m00}. All the radiation processes
are included in the calculations of the global accretion flow
structure. Integrating these equations from the outer boundary of the flow
at $R=R_{\rm out}$ inwards
the black hole, we can obtain the global structure of the accretion flow passing the
sonic point smoothly to the black hole horizon.
In our calculations, the values of some parameters adopted are different
from those in \citet{m00}, which will be discussed in Section 7. We can then calculate the spectrum
of the accretion flow based on this global structure of the RIAF.
In the spectral calculations, the gravitational redshift effect is considered, while the
relativistic optics near the black hole is neglected. We only calculate the total
luminosity radiated from the RIAF without considering its inclination. The derived
spectrum in this way can be taken as an average spectrum for AGNs, which
is a good approximation, as AGNs should have randomly distributed orientations.

\section{Spectra of standard disks}

For a standard thin disk, the flux due to viscous dissipation in
unit surface area  is \citep{ss73} \be F_{\rm vis}(R)\simeq {\frac {3GM_{\rm
bh}\dot M}{8\pi R^3}}\left[1-\left({\frac {3R_{\rm
S}}{R}}\right)^{1/2}\right], \label{fvis} \ee where $R_{\rm
S}=2GM_{\rm bh}/c^2$. The local disk temperature of the thin cold
disk is \be T_{\rm disk}(R)={\frac {F_{\rm
vis}^{1/4}(R)}{\sigma_{\rm B}^{1/4}}}, \label{tdisk} \ee by
assuming local blackbody emission. In order to calculate the disk
spectrum, we include an empirical color correction for the disk
thermal emission as a function of radius. The correction has the
form \citep{chiang02} \be f_{\rm col}(T_{\rm disk}) = f_\infty -
\frac{(f_\infty - 1) [1 +
                     \exp(-\nu_{\rm b}/\Delta\nu)]} { 1 +
                     \exp[(\nu_{\rm p} -\nu_{\rm b})/\Delta\nu]}, \label{fcol}
\ee where $\nu_p \equiv 2.82k_B T_{\rm disk}/h$ is the peak
frequency of blackbody emission with temperature $T_{\rm disk}$.
This expression for $f_{\rm col}$ goes from unity at low
temperatures to $f_\infty$ at high temperatures with a transition
at $\nu_{\rm b} \approx \nu_{\rm p}$. \citet{chiang02} found  that
$f_\infty = 2.3$ and $\nu_b = \Delta\nu = 5\times 10^{15}$\,Hz can
well reproduce the model disk spectra of \citet{h01}. The disk
spectra can therefore be calculated by \be L_\nu =32\pi^2 \left(
{\frac {GM_{\rm bh}}{c^2}} \right)^2 {\frac{h \nu^3}{c^2 } }
     \int\limits_{r_{\rm in}}^\infty
     {\frac{r dr}{f_{\rm col}^4[\exp(h\nu/f_{\rm col} k_B T_{\rm disk}) -
     1]}}, \label{sdspect}
\ee where $r_{\rm in}=R_{\rm in}/R_{\rm S}$ is the inner radius of
the standard disk. At a high accretion rate, $\dot{m}>\dot{m}_{\rm
crit}$, the standard thin disk extends to the minimum stable orbit
of the black hole, $r_{\rm in}=3$, for a non-rotating black hole. For a
RIAF+standard thin disk system, the spectrum emitted from the standard disk region
can be calculated by using the transition radius $r_{\rm tr}$
($r_{\rm tr}=R_{\rm tr}/R_{\rm S}$) instead of $r_{\rm in}$ as the
lower integral limit in Eq. (\ref{sdspect}).

\section{Transition radius $R_{\rm tr}$  between RIAF and
standard thin disk}

In this work, we assume the transition from a standard thin disk
to a RIAF to occur whenever $\dot{m}\la \dot{m}_{\rm crit}$, i.e.,
so-called "strong principle" \citep*[e.g.,][]{nmq98}. The RIAF is naturally expected to
match a standard thin disk at the transition radius $r_{\rm tr}$. The
detailed physics, causing such a transition of a standard thin
disk to a RIAF, is still unclear. It is suggested that the
standard thin disk transits is truncated at an initial
transition radius $r_{\rm tr,0}$ and a RIAF is present within this radius,
when $\dot{m}=\dot{m}_{\rm crit}$. The transition radius $r_{\rm tr}$ increases with
decreasing accretion rate $\dot{m}$ as \be r_{\rm
tr}\propto\dot{m}^{-p}, \label{dotmp} \ee where $p=2$ is predicted, based on the
scenario of transition triggered by the thermal instability, by
\citet{a95}; or $p\simeq0.8-1.3$ is expected by the disk
evaporation induced transition scenarios \citep{liu99,rc00,sd02}.
In either one of these transition scenarios, the transition radius
$r_{\rm tr}$ always increases with decreasing accretion rate
$\dot{m}$. The precise initial transition radius $r_{\rm tr,0}$ is still unknown, though it
should be small. In this work, we adopt $r_{\rm tr,0}=20$ in all our calculations.

\section{The observed distribution of the Eddington ratio $L_{\rm B}/L_{\rm Edd}$}

\begin{figure}
\centerline{\psfig{figure=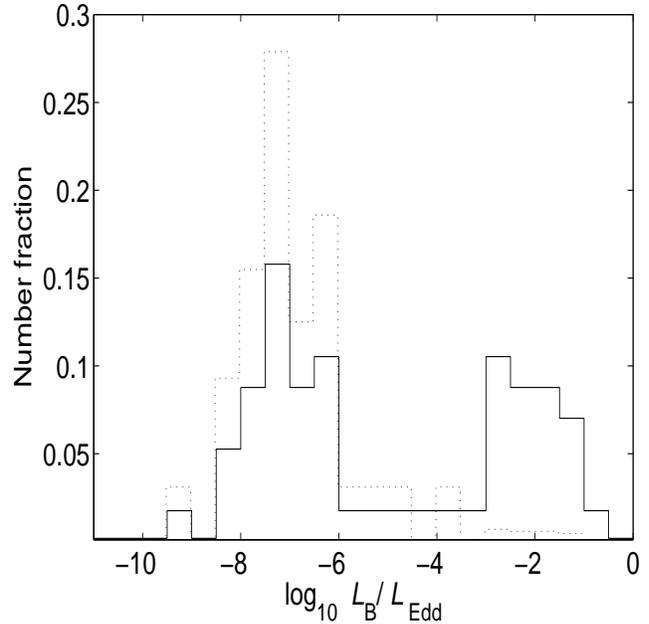,width=8.5cm,height=8.5cm}}
\caption{The distribution of the Eddington ratio $L_{\rm B}/L_{\rm Edd}$ at B-band
for the sample given by \citet{ho02} (the solid line). The dotted line is the
distribution for the effective
volume-corrected sample (see Section 5 for the details).
}
\label{fig1}
\end{figure}

In this work, we adopt the sample given by \citet{ho02}, in which
74 nearby supermassive black holes with both measured masses and B-band luminosities.
The black hole masses of the sources in this sample
have been measured by using two different approaches: stellar and gas kinematics and
reverberation mapping.  \citet{thp00} found that X-ray luminosities in 2-10 keV of LINERs (low-ionization nuclear
emission-line regions) with broad H$\alpha$ emission in their optical spectra
are proportional to their H$\alpha$ luminosities. This indicates
that the dominant ionizing source in LINERs is photoionization by hard photons
from low-luminosity AGNs. The B-band luminosities of the sources in
this sample are estimated from the line emissions, which
are supposed to be photo-ionized by the nuclear radiations. This may cause some uncertainties
for individual sources, but the derived Eddington ratio should still be reliable in statistical
sense. This sample includes 17 PG quasars, which may not be in the same population with those
low-luminosity counterparts and are accreting at higher rates $\dot{m}>\dot{m}_{\rm crit}$.
As we are focusing on the inefficient accretion history of AGNs, we leave out these 17
PG quasars, which leads to 57 sources. The Eddington ratio distribution
for this sample is plotted in Fig. \ref{fig1}. This sample was used to explore the evolution of low-luminosity AGNs by \citet{hnh06}.
The selection effects of this sample were extensively analyzed in their work (see Section 2.1
in their paper for the details). They found that, the FR I radio galxies from the sample used
by \citet{mcf04} exhibit significant different luminosity distribution from that of the
sample by \citet{ho02}, while both of these two sample exhibit similar
Eddington ratio distributions. This means that the results of AGN evolution derived from either
one of these two samples would be qualitatively unchanged, which may imply that the derived
results are not affected significantly by selection effects.

The maximal distance for the sources measured by reverberation mapping method
is about 3 times that of the subsample measured by the stellar/gas kinematics.
\citet{hnh06} suggested to multiply the relative fraction of the sources measured kinematics
by 3$^3$ to account for the different subsample volumes. The derived distribution for this
tentatively volume-corrected sample is also plotted in Fig. \ref{fig1}.

\section{The time-dependent accretion rate $\dot{m}(t)$ derived from the Eddington
ratio distribution}

The light curve of AGNs at B-band can be derived from the observed Eddington ratio distribution
${\rm d}N/{\rm d}\log (L_{\rm B}/L_{\rm Edd})$:
\begin{equation}
{\frac {{\rm d}\log (L_{\rm B}/L_{\rm Edd})}{{\rm d}t}}=N^{\rm tot}
\left( {\frac {{\rm d}N}{{\rm d}\log (L_{\rm B}/L_{\rm Edd})} }\right)^{-1}, \label{lcurve}
\end{equation}
where $N^{\rm tot}$ is the total number of the sources in the sample, and $t$ is normalized to unity.
The time-dependent accretion rate $\dot{m}(t)$ can be derived from the light curve by
\begin{equation}
{\frac {{\rm d}\dot{m}(t)}{{\rm d}t}}={\frac {{\rm d}\log (L_{\rm B}/L_{\rm Edd})}{{\rm d}t}}
\left({\frac {{\rm d}\log (L_{\rm B}/L_{\rm Edd})}{{\rm d}\dot{m}}}\right)^{-1},\label{mdott}
\end{equation}
where ${\rm d}\log (L_{\rm B}/L_{\rm Edd})/{\rm d}\dot{m}$ is available based on spectral
calculations for the accretion mode transition model desribed in Sections 2-4.
Here, we have to assume that all sources have the same time-dependent accretion rate $\dot{m}(t)$,
and it evolves monotonically with time. The latter assumption may not be the case for
some individual sources in short timescales, but it should be reseasonable in statistic
sense for a sample of AGNs, because accretion rates should decline in a long timescale. 
Substituting Eq. (\ref{lcurve}) into Eq. (\ref{mdott}), we obtain
\begin{equation}
{\frac {{\rm d}\dot{m}(t)}{{\rm d}t}}=N^{\rm tot}
\left({\frac {{\rm d}N}{{\rm d}\dot{m}}}\right)^{-1}. \label{Nmdot}
\end{equation}
Using the X-ray luminosity function of AGNs given by \citet{u03}, we calculate
number density of AGNs in comoving space as a function of redshift. It is found that
the number density of AGNs in the local universe is about 45 per cent of that at $z=0.5$, which
implies that the AGNs at low redshifts are roughly in a steady evolving state with switching on
and off being in balance. The cosmological evolution of AGNs at low redshifts should be
unimportant, which will affect our results very little, as the sample used in this work
is limited to the sources in the local universe.

\begin{figure}
\centerline{\psfig{figure=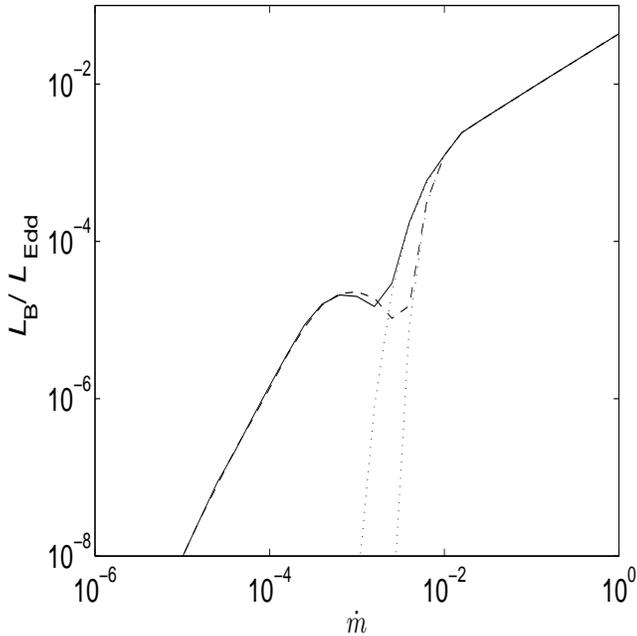,width=8.5cm,height=8.5cm}}
\caption{The spectral evolution of RIAF$+$standard thin disk systems with varying
transition radii $r_{\rm tr}$. The dotted lines represent the
emission from the outer standard thin disk regions. The solid line
represents the case for $p=1$, while the dashed line is for $p=2$.   } \label{fig2}
\end{figure}

\begin{figure}
\centerline{\psfig{figure=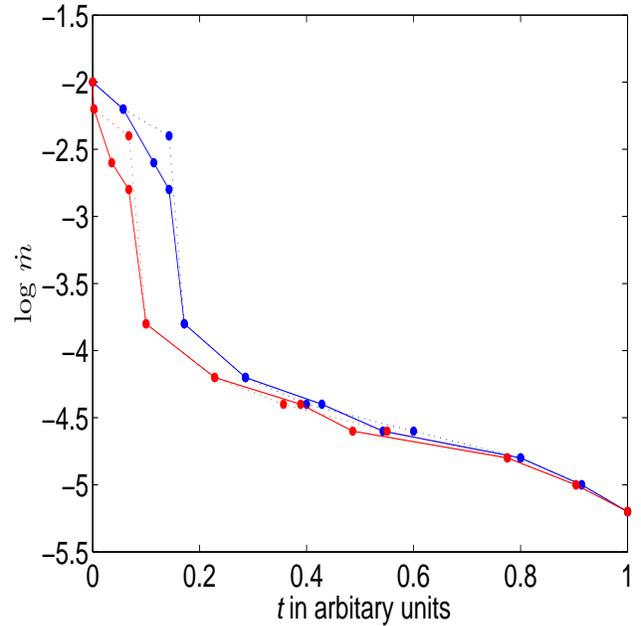,width=8.5cm,height=8.5cm}}
\caption{The derived time-dependent accretion rates $\dot{m}(t)$. The blue lines
represent the cases derived directly from \citet{ho02}'s sample, while the red lines
are for the effective volume-corrected sample. The solid lines are for the cases of
$p=1$, while the dotted lines are for $p=2$. } \label{fig3}
\end{figure}

\section{Results}

The detailed physics of the viscosity in accretion disks is still
quite unclear, and it is usually described by the viscosity
parameter $\alpha$. Assuming accretion to be driven by turbulent
stresses generated by the magnetorotational instability, the
three-dimensional MHD simulations suggest that the viscosity
parameter $\alpha$ in the accretion flows is $\sim 0.1$
\citep{a98}, or $\sim 0.05-0.2$ \citep{hb02}. The critical
accretion rate $\dot{m}_{\rm crit}\simeq 0.01$ for accretion mode
transition is suggested by different authors either from
observations or theoretical arguements \citep*[see,
e.g.,][]{nmq98}. Our numerical calculations for the global
structure of the flows show $\dot{m}_{\rm crit}\simeq 0.01$ for
$\alpha=0.2$. In this work, we adopt the $\alpha$-viscosity,
$\alpha=0.2$, and limit our calculations of the RIAF structure to
hydrodynamics, but the magnetic pressure is included by a
parameter $\beta$. The parameter $\beta$, defined as $p_{\rm
m}=(1-\beta)p_{\rm tot}$, describes the magnetic field strength of
the gases in the flow. This parameter is in fact not a free
parameter, which is related to the viscosity parameter $\alpha$ as
$\beta\simeq (6\alpha-3)/(4\alpha-3)$, as suggested by the MHD
simulations \citep{hgb96,nmq98}. For $\alpha=0.2$, $\beta\simeq
0.8$ is required. \citet{m00} adopted very small values of
$\delta$ in the calculations, as those of traditional advection
dominated accretion flow (ADAF) models \citep*[see][for a review,
and references therein]{narayan02}. It was pointed out that such a
small $\delta$ adopted in traditional ADAF models is very
unlikely, because a significant fraction of the viscously
dissipated energy (up to $\delta \sim 0.5$) could go into
electrons by magnetic reconnection, if the magnetic fields in the
flow are strong \citep{bl97,bl00}. The average black hole mass for
the sources in this sample is $\simeq 2.4\times 10^8 {\rm
M}_\odot$. In our calculations, we adopt $\delta=0.5$, and $M_{\rm
bh}=2.4\times 10^8 {\rm M}_\odot$. Our numerical calculations show
that the dimensionless light curves $L_{\rm B}/L_{\rm Edd}$ as
function of $\dot{m}$ change very little for different black hole
masses adopted (in the range of $\sim 10^{6-10} {\rm M}_\odot$).

Using the method described in Sections 2 and 3, we can calculate
the global accretion flow structure and then the spectrum varying
with accretion rate $\dot{m}$ and $r_{\rm tr}$. The initial
transition radius $r_{\rm tr,0}=20$ is adopted in all our
calculations for RIAFs. The Eddington ratios $L_{\rm B}/L_{\rm
Edd}$ evolving with $\dot{m}$ are plotted in Fig. \ref{fig2}. The
time-dependent accretion rates $\dot{m}(t)$ can be derived from
the observed Eddington ratio distributions by using Eqs.
(\ref{lcurve}) and (\ref{mdott}) (see Fig. \ref{fig3}).

\section{Discussion}

For a RIAF$+$standard thin disk system, its optical
emission is mostly from the outer thin accretion disk
region, if its accretion rate $\dot{m}$ is close to the critical
value $\dot{m}_{\rm crit}$, because the transition radius is small
for this case. For a large transition radius, corresponding to a low
$\dot{m}$, the temperature of the
standard thin disk region is very low, and the optical emission is
dominantly from the inner RIAF (see Fig. \ref{fig2}).
The optical luminosity drops rapidly with $\dot{m}$, when the flow is
accreting at $\la \dot{m}_{\rm crit}$.

In Fig. \ref{fig2}, we find that the luminosity decreases rapidly
after the accretion mode transition at $\dot{m}=\dot{m}_{\rm crit}\sim 0.01$.
It is found that the light curves $L_{\rm B}(\dot{m})$ are almost same for $p=1$ and $p=2$,
except the accretion rate $\dot{m}$ is in the range of $\sim 0.004-0.01$. When the accretion rate $\dot{m}$ is slightly lower than $\dot{m}_{\rm crit}$,
the emission is dominantly from the outer thin disk regions, because the standard thin
disks still extend to small radii. The emission becomes dominated by the
radiation from the inner RIAFs, when the accretion rate $\dot{m}\la 5\times 10^{-3}$.

Based on the relations between accretion rate $\dot{m}$ and B-band luminosity
$L_{\rm B}(\dot{m})$ given in
Fig. \ref{fig2}, we can derive the time-dependent accretion rates $\dot{m}(t)$ from
the observed Eddington ratio distributions for AGNs. In Fig. \ref{fig3}, we plot the
derived $\dot{m}(t)$ for the cases with two different values of $p$ for variable transition
radius models, from the Eddington ratio distribution of the \citet{ho02}'s sample
or the sample with effective volume-corrected (see Section 5), respectively.
We find that there is a rapid decrease of $\dot{m}(t)$ from $10^{-2}$ to $\sim 10^{-4}$
either for the cases with different values of $p$ or from
different Eddington ratio distributions (the original one or the effective volume-corrected
one).  It is not surprised to find that the derived time-dependent accretion rates $\dot{m}(t)$
are quite similar for the cases with $p=1$ or $p=2$, because the theoretical curves
$L_{\rm B}(\dot{m})$ are very similar for these two different values of $p$ (see Fig. \ref{fig2}).
For the Eddington ratio distribution of the effective volume-corrected sample, there are
more sources with low luminosities, i.e., less sources with $\dot{m}$ close to
$\dot{m}_{\rm crit}$, so the derived time-dependent $\dot{m}(t)$ decreases more
rapidly than that derived from the original sample. We find that the main feature of
the time-dependent $\dot{m}(t)$ has not been changed by this effective volume-corrected sample.
\citet{cao05} calculated the hard
X-ray emission from all RIAFs in faint AGNs, and compared them with the observed X-ray
background. It was found that the accretion rate should decrease rapidly to the value far
below the critical rate within a timescale shorter than 5 per cent of bright quasars'
lifetime, which is consistent with our present time-dependent accretion rates $\dot{m}(t)$
derived from the Eddington ratio distributions. The present derived time-dependent 
$\dot{m}(t)$ is based on the assumption of monotonical evolution with time. 
From Eq. (\ref{Nmdot}), we find that ${\rm d}\dot{m}(t)/{\rm d}t$ 
also represents the distribution of accretion rates $\dot{m}$ for the sample. 

The present sample may have overlooked a number of sources with very low Eddington ratios
($\la 10^{-5}-10^{-6}$), which means a complete sample should include more
sources accreting at very low rates $\la 10^{-4}$ than the present sample. This implies
that the relative timescale of the rapid drop of accretion rate from $0.01$ to $\sim 10^{-4}$
should be even shorter than those plotted in Fig. \ref{fig3}, if a complete sample is adopted.
Such a complete sample is still unavailable now, however, the main feature of $\dot{m}(t)$
with a rapid declining between $0.01$ and $\sim 10^{-4}$ derived in this work will not
be changed qualitatively.

The fraction of the viscously dissipated energy that directly goes
into electrons, $\delta=0.5$, is adopted in this work, as suggested by \citet{bl97}. The precise
value of $\delta$ is still unknown, and it may slightly be lower than $0.5$. If a slightly
lower $\delta$ is adopted, the derived light curves $L_{\rm B}(\dot{m})$ have the similar form,
but they should be systematically lower than those in Fig. \ref{fig2}. Thus, the main
feature of $\dot{m}(t)$, a rapid decrease after the accretion mode transition, will still be
present as those plotted in Fig. \ref{fig3}.

The RIAF may have winds, and a power-law $r$-dependent accretion rate is assumed,
though the detailed physics is still unclear \citep{bb99}. When the accretion rate
$\dot{m}$ is close to 0.01, the optical emission is dominated by that from the outer
standard thin disk region (see Fig. \ref{fig2}). Thus, the optical spectrum of the
RIAF+standard thin disk system will not be affected by the winds of the RIAF, when
$\dot{m}$ is close to 0.01.
The optical emission from the RIAFs is attributed to Compton up-scatterings of the soft
synchrotron photons by the hot electrons in the accretion flows.
Our numerical calculations on the RIAF spectra show that most optical emission (more
than 90 per cent) is from the regions within 10 $R_{\rm S}$. This implies that
the optical emission of a RIAF is similar to a RIAF with winds, provided they have the
same accretion rate at their inner edge. In this
work, we focus on the accretion rate at the inner edge of the accretion flow, i.e., the
mass of gases swallowed by the black hole.
Thus, the derived time-dependent accretion rates at the inner edges of the RIAFs will not
be altered much, even if winds are present in the RIAFs.

In all our calculations. the initial transition radius $r_{\rm tr,0}=20$ is adopted.
As the transition radius $r_{\rm tr}$ increases rapidly with decreasing accretion rate
$\dot{m}$ (see Eq. \ref{dotmp}), the resulted light curves $L_{\rm B}(\dot{m})$ will only be
slightly different for different values of $r_{\rm tr,0}$ adopted, provided $r_{\rm tr,0}$
is not very large. Thus, the derived $\dot{m}(t)$ is insensitive to the value of $r_{\rm tr,0}$.

In principle, one has to consider the activity trigger rates of galaxies along the cosmic time
in the study of AGN evolution. However, it is difficult to have a complete sample including both bright and
faint AGNs, which can be used to explore the evolution of bright quasars to faint AGNs.
In this work, we only explore the time-dependent $\dot{m}(t)$ for $\dot{m}\la\dot{m}_{\rm crit}$.
and the sources in the sample we used are limited in the local universe. Thus, the results may
not be affected by the unknown trigger rates, though it is unclear if
the derived $\dot{m}(t)$ in this work is valid for the AGNs at high redshifts.

\section*{Acknowledgments}
We thank A. Loeb for the explanation on their
quasar evolution models, R. Narayan for helpful conversation, and
the anonymous referees for their comments and suggestions.
This work is supported by the National Science Fund for
Distinguished Young Scholars (grant 10325314), and NSFC (grant
10333020).

\clearpage


\end{document}